\documentstyle[aps,pre,epsf,twocolumn,floats,fancyheadings,twoside]{revtex}
\begin{document}
\renewcommand{\arraystretch}{1.5}
\bibliographystyle{prsty}
\date{\today}
\twocolumn[\hsize\textwidth\columnwidth\hsize\csname
@twocolumnfalse\endcsname

\title{\bf Order Parameter Equations for Front Transitions: 
           Planar and Circular Fronts}

\author{A. Hagberg\thanks{\tt aric@lanl.gov }}
\address{Center for Nonlinear Studies and T-7,\\ Theoretical Division,
         Los Alamos National Laboratory, Los Alamos, NM 87545}

\author{E. Meron\thanks{\tt ehud@bgumail.bgu.ac.il}}
\address{The Jacob Blaustein Institute for Desert Research and
         the Physics Department, \\ Ben-Gurion University, 
         Sede Boker Campus 84990, Israel}

\author{I. Rubinstein and B. Zaltzman}
\address{The Jacob Blaustein Institute for Desert Research and
        the Mathematics Department, \\Ben-Gurion University, 
        Sede Boker Campus 84990, Israel}

\maketitle

\begin{abstract}
Near a parity breaking front bifurcation, small perturbations may
reverse the propagation direction of fronts.  Often this results in
nonsteady asymptotic motion such as breathing and domain breakup.
Exploiting the time scale differences of an activator-inhibitor
model and the proximity to the front bifurcation, we derive
equations of motion for planar and circular fronts.  The equations involve a
translational degree of freedom and an order parameter describing
transitions between left and right propagating fronts.  
Perturbations, such as a space dependent
advective field or uniform curvature (axisymmetric spots),
couple these two degrees of freedom.  In both cases this
leads to a transition from stationary to oscillating
fronts as the parity breaking bifurcation is approached.
For axisymmetric spots, two additional dynamic behaviors are found: 
rebound and collapse. 
\end{abstract}


\vskip2pc]
\narrowtext
%
%
\section{Introduction}

Pattern dynamics in reaction-diffusion systems often involve nonsteady
front motions.  These motions can be driven by 
curvature~\cite{TyKe:88,Meron:92}, 
front interactions~\cite{EMS:88,EMS:90}, convective
instabilities~\cite{AKP:84,MPH:88}, and external 
fields~\cite{Ort:87,SSM:92,MHM:94}. 
In some cases fronts reverse their direction of 
propagation, as for example in 
breathing pulses~\cite{KoKu:80,KeOS:82,NiMi:89,OITe:90,RRHM:93,HaMe:94a,SOMS:95,MuOs:96,WSBOP:96},
where the reversal is periodic
in time, and nucleation of spiral-vortex pairs, where the reversal is
local along the extended front
line~\cite{HaMe:94b,EHM:95,LMOS:93,LMSP:94,LeSw:95}.

Earlier studies demonstrated that front reversals, as described above,
become feasible near a nonequilibrium Ising-Bloch
(NIB) bifurcation~\cite{HaMe:94a,CLHL:90}, that is, a parity breaking 
bifurcation where a single stationary front loses stability 
to a pair of new, counterpropagating fronts. 
The reversal phenomenon can be regarded as a dynamic transition
between the left and right propagating fronts 
that appear beyond the front bifurcation. It is induced by 
intrinsic perturbations,
like curvature and front interactions~\cite{HaMe:94b,HaMe:94c}, or weak 
space dependent external fields~\cite{EHM:95}. Since the left and right 
propagating fronts differ in internal structure~\cite{HaMe:94a}
such a transition involves a new degree of freedom in addition to
the translation mode: the order parameter associated with the
bifurcation~\cite{EHM:95}.  The effect of the perturbations is to
couple these two degrees of freedom in a way that allows for front reversal.

Our objective is to derive equations for front motion 
that capture front reversal.
Progress toward that goal has already been made in
Ref.~\cite{EHM:95,Bode:96} for a nondiffusing inhibitor.  Since inhibitor
diffusion is essential for spontaneous front reversals induced by
curvature, in this work we study the more difficult case of a
diffusing inhibitor. 
This calls for a different approach as described in
Sections~\ref{formfree} and~\ref{solfree}.
In Section~\ref{sec:reversal} we study front reversals induced by
two types of perturbations of planar fronts, an external advective
field and uniform curvature.
In this case only planar and circular fronts are studied; the
derivation of the more general equations for nonuniformly curved fronts
can be found in Ref.~\cite{HaMe:97}.
Some of the results
presented here have been briefly reported in Ref.~\cite{HMRZ:96}.

%
%
\section{Reaction-diffusion model}
The model we consider is an activator-inhibitor reaction-diffusion
system describing a bistable medium.  Models of this type have been
studied in various physical and chemical 
contexts~\cite{TyKe:88,Meron:92,BNREG:95,HBKR:95,BRSP:94,SBP:95}. 
The specific form chosen here is
\begin{eqnarray}
\label{pde1}
u_t&=&\epsilon^{-1}(u-u^3-v)+\delta^{-1} u_{xx}\,, \\
v_t&=&u-a_1v+ v_{xx} \,. \nonumber
\end{eqnarray}
The variables $u$ and $v$ are scalar real fields representing the
activator and the inhibitor, respectively, with the subscripts $x$ and $t$
denoting partial derivatives with respect to these variables.  For $a_1>1$
the system~(\ref{pde1}) has two stationary uniform states,
$(u_\pm,v_\pm)=
\left(\pm\sqrt{1-1/a_1},\pm a_1^{-1}\sqrt{1-1/a_1}\right)$. Note
that the parity symmetry $(u,v)\to (-u,-v)$ of Eqs.~(\ref{pde1}) 
is reflected in these solutions. Generalizations of Eqs.~(\ref{pde1}) 
to nonsymmetric forms in one and two space dimensions will be
considered in Section~\ref{sec:reversal}.

\begin{figure}
\epsfxsize=3.5 truein \centerline{\epsffile{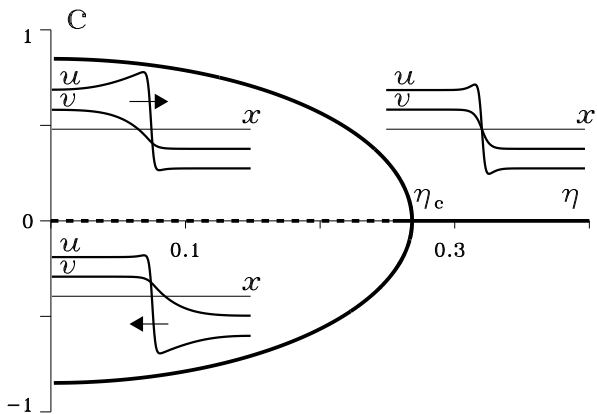}}
\caption{The NIB (or nonequilibrium Ising-Bloch) bifurcation and internal
structure of front solutions.  The pitchfork diagram represents
the speed of front solutions $vs$ the parameter $\eta$.
For $\eta>\eta_c$, the Ising front
is the single solution and $v_f$, the order parameter representing the value of
the $v$ field at the front position $u=0$, is zero.  Beyond the
bifurcation, $\eta<\eta_c$, a pair of counterpropagating
Bloch fronts appears.  The order parameter $v_f$ is negative 
(positive) for rightward (leftward) propagating fronts.}
\label{fig:fronts}
\end{figure}

In addition to the spatially uniform solutions there are also
front solutions connecting them.  In the following we will
consider front solutions that connect $(u_+,v_+)$ at $x=-\infty$
to $(u_-,v_-)$ at $x=\infty$.  The number and type of these front
solutions is determined by the two parameters $\epsilon$ and
$\delta$.  For $\eta>\eta_c=3/ 2\sqrt2 q^3$, where
$\eta=\sqrt{\epsilon\delta}$ and $q^2=a_1+1/2$, a
single stable stationary (Ising) front solution exists.  This
solution loses stability in a pitchfork bifurcation, at
$\eta=\eta_c$, to a pair of counterpropagating (Bloch) front
solutions~\cite{HaMe:94a,BRSP:94,SBP:95,IMN:89} as
shown in Fig.~\ref{fig:fronts}. 
The two Bloch front solutions differ
not only in their propagation direction but in their internal
structure.  
This difference can be represented by an order
parameter associated with the bifurcation, which for 
$\mu=\epsilon/\delta\ll 1$ can be taken to be the value $v_f$ of
the $v$ field at the front position.  For simplicity we define
the front position to be at $u=0$. With this choice, $v_f$=0 for
the Ising front 
(the inset for $\eta>\eta_c$ in Fig.~\ref{fig:fronts}).  Beyond the
front bifurcation, $v_f$ is nonzero and the sign indicates the
direction of front propagation: $v_f<0$ for the front propagating
to the right and $v_f>0$ for the front propagating to the left
($\eta<\eta_c$ in Fig.~\ref{fig:fronts}).  

%
%
\section{Formulation of the free boundary problem}
\label{formfree}

In the following we confine ourselves to the region
$\epsilon\ll 1$, $\delta\propto\epsilon^{-1}$  
and we choose $\delta$ values such 
that $\epsilon\delta\sim{\cal O}(\eta_c^2)$.
The small parameter $\epsilon$ allows the use of singular
perturbation methods to study front solutions to Eqs.~(\ref{pde1}).
The first step is to transform to a moving coordinate frame, 
$x\to r=x-x_f(t)$, $t\to t$ where $x_f$ is the position of the front.
In this frame Eqs.~(\ref{pde1}) become
\begin{eqnarray}
\label{mframe}
u_t-\dot x_fu_r&=&\epsilon^{-1}(u-u^3-v)+\delta^{-1}u_{rr}\,, \\
v_t-\dot x_fv_r&=&u-a_1v+v_{rr} \nonumber \,,
\end{eqnarray}
where the dot over $x_f$ denotes the derivative with respect to $t$. The 
front solution, $u(r,t)$, $v(r,t)$, is characterized
by a strong variation of the $u$ field over a distance of order 
$\sqrt\mu=\sqrt{\epsilon/\delta}$.  Stretching the spatial 
coordinate, $z=r/\sqrt\mu$, to expand this region
Eqs.~(\ref{mframe}) become
\begin{eqnarray}
\label{inner}
\epsilon\,(u_t-\dot z_fu_z)&=&u-u^3-v+u_{zz}\,, \\
\mu\,(v_t-\dot z_fv_z-u+a_1v)&=&v_{zz}\nonumber \,,
\end{eqnarray}
where $z_f=x_f/\sqrt\mu$ and we recall that $\mu\propto\epsilon^2$.  
Expanding in $\epsilon$
\begin{eqnarray}
u&=&u_0+\epsilon u_1 + \epsilon^2 u_2 +... \,,\nonumber \\
v&=&v_0+\epsilon v_1 + \epsilon^2 v_2 +... \,,\nonumber
\end{eqnarray}
and inserting into~(\ref{inner}) we find at order unity 
the front solution
\[ u_0=-\tanh(z/\sqrt 2),\qquad v_0=0\,. \]

Collecting terms of order $\epsilon$ gives
\begin{equation}
{\cal L}  u_1= v_1-\dot z_f u_{0z}\,,\qquad{\cal L}=
\partial_z^2+1-3u_0^2 \,,\label{lin}
\end{equation}
where $v_1$ is a yet undetermined function of time.
Since 
\[{\cal L}u_{0z}=0\,,\]
solvability of~(\ref{lin}) requires
\[ \dot z_f =-{3\over \sqrt 2} v_1(t) \,. \]
The narrow front region becomes
infinitely thin in the limit $\epsilon\to 0$.  Therefore, 
$v(t)$ may be associated with the value of $v(r,t)$ at the front position, 
that is $v(0,t)$.
With this notation the leading order relation is
\begin{equation}
\dot x_f =-{3\over\eta \sqrt 2} v(0,t)\,.\label{pos}
\end{equation}

Away from $x_f$,  $u-u^3-v\sim {\cal O}(\epsilon)$,
and $u$ varies on the same time and length scales as $v$. Going back to
Eqs.~(\ref{mframe}) we find at order unity
\begin{eqnarray}
\label{outer}    
v_t-\dot x_fv_r=u_+(v)-a_1v+v_{rr}\,,&\qquad&r\le 0\,,\\   
v_t-\dot x_fv_r=u_-(v)-a_1v+v_{rr}\,,&\qquad&r\ge 0\,, \nonumber
\end{eqnarray}
where $u_\pm(v)$ are the outer solution branches of the cubic equation
$u-u^3-v=0$. For $a_1$ sufficiently large we may linearize the branches 
$u_\pm(v)$ around $v=0$
\begin{equation}
u_\pm(v)\approx \pm 1-v/2\,.\label{linap}
\end{equation}
Substituting the linearization~(\ref{linap}) and the relation from the
inner problem~(\ref{pos}) into~(\ref{outer}) produces the free 
boundary problem
\begin{equation}
\begin{array}{rl}
\left.
\begin{array}{rcl}
v_t+q^2v-v_{rr}&=&1- {\displaystyle \frac{3}{\eta\sqrt 2}}v(0,t)v_r \\
v(-\infty,t)&=&v_+\approx q^{-2}  \\
\end{array}
\quad \right\}
& \quad r\le 0\,, \\
\\
\left.
\begin{array}{rcl}
v_t+q^2v-v_{rr}&=&-1-{\displaystyle\frac{3}{\eta\sqrt 2}}v(0,t)v_r \\
v(\infty,t)&=&v_-\approx -q^{-2}  \\
\end{array}
\quad \right\}
& \quad r\ge 0\,, 
\end{array}
\label{free}
\end{equation}
\begin{equation}
\left[v\right]_{r=0}=\left[v_r\right]_{r=0}=0\,,\label{bracket}
\end{equation}
where the square brackets in~(\ref{bracket}) denote jumps across the free 
boundary.  
The solution to~(\ref{free}) leads to 
a dynamic equation for $v_f(t)=v(0,t)$, 
the value of the inhibitor at the front position $x_f(t)$, which will 
complement Eq.~(\ref{pos}).

%
%
\section{Solution of the free boundary problem}
\label{solfree}

Near the front bifurcation ($\eta$ close to $\eta_c$), 
the front speed, $c$, is small and propagating front
solutions can be expanded as power series in $c$ around the stationary
front solution.  The stationary front solution satisfies the boundary value 
problem
\begin{equation}
\begin{array}{rl}
\left.
\begin{array}{rcl}
v_{rr}-q^2v+1&=&0 \\
v(-\infty)&=&q^{-2}\\
v(0)&=&0 \\
\end{array}
\qquad
\right\}
& \qquad r\le 0\,, \\
\\
\left.
\begin{array}{rcl}
v_{rr}-q^2v-1&=&0 \\
v(\infty)&=&-q^{-2} \\
v(0)&=&0  \\
\end{array}
\qquad
\right\}
& \qquad r\ge 0\,, 
\end{array}
\label{stat}
\end{equation}
with $\left[v_r\right]_{r=0}=0$. The solution to~(\ref{stat}) is
\begin{eqnarray}
\label{statsol}
v^{(0)}(r)&=q^{-2}(1-e^{qr})\,,& \qquad r\le 0\,, \\
v^{(0)}(r)&=q^{-2}(e^{-qr}-1)\,,& \qquad r\ge 0\,. \nonumber
\end{eqnarray}
Note that $v^{(0)}_r=-q^{-1}\exp(-q\vert r\vert)$ 
is not differentiable at $r=0$.

In terms of the deviation from the stationary solution,
$\bar v=v-v^{(0)}$, Eqs.~(\ref{free}) are
\begin{eqnarray}
\label{free1}
\bar v_t+q^2\bar v-\bar v_{rr}&=&
-{3\over\eta \sqrt2}v(0,t)(\bar v_r+v^{(0)}_r)\,,\\
\bar v(\pm\infty)&=&0\,,\nonumber
\end{eqnarray}
We seek propagating solutions of~(\ref{free1}) that involve two time
scales, the 
original time $t$ and a slow time $T=c^2 t$. The slow time dependence is
a result of the slow evolution of $v_f$ (the value of the inhibitor at the
front position) near the front bifurcation. It is 
easy to show that to linear order $\dot v_f\propto (\eta_c-\eta)v_f$ and 
for a pitchfork bifurcation $\eta_c-\eta\propto c^2$, hence the $c^2$ scale. 
Expanding $\bar v(r,t,T)$ in powers of $c$ and $\eta$ in powers of $c^2$
(expecting a pitchfork bifurcation),
\begin{equation}
\bar v(r,t,T)=\sum_{n=1}^\infty c^n v^{(n)}(r,t,T)\,,\label{vbar}
\end{equation}
\begin{equation}
\eta=\eta_c-c^2\eta_1+c^4\eta_2+...\,,\label{eta}
\end{equation}
and inserting in~(\ref{free1}) gives the sequence of equations
\begin{equation}
v^{(n)}_t+q^2v^{(n)}-v^{(n)}_{rr}=-\rho^{(n)}\,,
\qquad n=1,2,3,\label{prob}
\end{equation}
where
\begin{mathletters}
\label{rho}
\begin{eqnarray}
\rho^{(1)}&=&
{3\over\sqrt2\eta_c}v^{(1)}_{\vert r=0}v^{(0)}_r\,,\label{rho1}\\
\rho^{(2)}&=&
{3\over\sqrt2\eta_c}
\left[v^{(1)}_{\vert r=0}v^{(1)}_r+
v^{(2)}_{\vert r=0}v^{(0)}_r\right]\,, \label{rho2}\\  
\rho^{(3)}&=&
v^{(1)}_T+{3\eta_1\over\sqrt 
2\eta_c^2}v^{(1)}_{\vert r=0}v^{(0)}_r \label{rho3} \\
&&\mbox{}+{3\over\sqrt2\eta_c}
\left[v^{(1)}_{\vert r=0}v^{(2)}_r
+v^{(2)}_{\vert r=0}v^{(1)}_r 
+v^{(3)}_{\vert r=0}v^{(0)}_r\right] \nonumber \,.
\end{eqnarray}
\end{mathletters}

Equation~(\ref{prob}) can be solved using an appropriate Green's function and
assuming that the front motion is 
independent of the fast time scale $t$ as $t\to\infty$~\cite{HMRZ:96}. 
A simplified derivation
of the solution follows from the gradient nature of (\ref{prob}) when the 
source term $\rho^{(n)}(r,T)$ becomes independent of $t$. For then
$v^{(n)}_t\to 0$ as $t\to\infty$ for any $r$ and we can
look for stationary ($t$ independent) solutions.

Consider first $v^{(1)}$. Inserting~(\ref{rho1}) in  
$v^{(1)}_{rr}-q^2v^{(1)}=\rho^{(1)}$ and solving for $v^{(1)}$ we obtain
\begin{equation}
v^{(1)}(r,T)={3\over 2\sqrt 2\eta_c q^3}v^{(1)}(0,T)F(r)\,,\label{fun1}
\end{equation}
where
\begin{eqnarray}
F(r)&=(1-qr)e^{qr}\,,& \qquad r\le 0\,, \nonumber \\
F(r)&=(1+qr)e^{-qr}\,,& \qquad r\ge 0\,. \nonumber
\end{eqnarray}
Setting $r=0$ we find
\begin{equation}
\eta_c={3\over 2\sqrt 2 q^3}\,.\label{comp1}
\end{equation}
The critical value $\eta_c=\eta(c=0)$ determines the bifurcation point 
where the propagating Bloch front solutions coincide with the stationary Ising
front. The expression~(\ref{comp1}) is the same as the one derived earlier 
using a different method~\cite{HaMe:94a,HaMe:94c}.     

Using~(\ref{fun1}) to solve 
$v^{(2)}_{rr}-q^2v^{(2)}=\rho^{(2)}$ we find
\begin{equation}
v^{(2)}(r,T)=\left(v^{(2)}(0,T) + {1\over 2}{v^{(1)}(0,T)}^2 q^3 r\right)
F(r)\,,\label{fun2}
\end{equation} 
and using both~(\ref{fun1}) and~(\ref{fun2}) in
$v^{(3)}_{rr}-q^2v^{(3)}=\rho^{(3)}$ we obtain
\begin{equation}
\begin{array}{rl}
\left.
\begin{array}{rcl}
v^{(3)}(r,T)&=&v^{(3)}(0,T)e^{qr}+A_+ re^{qr} \\
&&\mbox{}- B_+ r^2e^{qr}-C_+ r^3 e^{qr} \,, \nonumber 
\end{array}
\right\}
& r\le 0\,, \\
\\
\left.
\begin{array}{rcl}
v^{(3)}(r,T)&=&v^{(3)}(0,T)e^{-qr} + A_- re^{-qr} \\
&&\mbox{} - B_- r^2e^{-qr} -C_- r^3 e^{-qr}  \,, \nonumber
\end{array}
\right\}
& r\ge 0\,, 
\end{array}
\label{fun3}
\end{equation}
where
\begin{eqnarray}
A_\pm &=&q^3v^{(1)}(0,T)v^{(2)}(0,T) \nonumber \\
&&\mbox{}\pm\left[{3\over 4q}v^{(1)}_T(0,T)+{1\over 2}q^5 v^{(1)}(0,T)^3
\right.\nonumber \\
&&\mbox{}-\left.{q\eta_1\over\eta_c}v^{(1)}(0,T) - qv^{(3)}(0,T)\right]\,,  
\nonumber\\
B_\pm&=&{1\over 4}v^{(1)}_T(0,T) \pm q^4v^{(1)}(0,T)v^{(2)}(0,T)\,, 
\nonumber \\
C_\pm&=&\pm {1\over 6}q^7v^{(1)}(0,T)^3\,. \nonumber
\end{eqnarray}
Application of the (no) jump condition $\bigl[v^{(3)}_r\bigr]_{r=0}=0$ leads
to 
\begin{equation}
\eta_c^2 v^{(1)}_T(0,T)={\sqrt 2\eta_1\over q} v^{(1)}(0,T)-{3\over 
4}{v^{(1)}}(0,T)^3 \,. \label{comp2}
\end{equation}

Equation~(\ref{comp2}) still contains an unspecified parameter, $\eta_1$. 
To identify
$\eta_1$ recall that $c$ is the speed of a front propagating at constant 
velocity.  From~(\ref{pos})
\[ \vert\dot x_f\vert={3\over\sqrt 2\eta_c}c\vert 
v^{(1)}(0,T)\vert+{\cal O}(c^2)\,.\]
Identifying $\vert\dot x_f\vert$ with $c$ gives
${v^{(1)}}^2=2\eta_c^2/9$
for a front propagating at constant speed. This
value of ${v^{(1)}}^2$ should coincide with the nontrivial
stationary solution of~(\ref{comp2}), ${v^{(1)}}^2=4\sqrt 2\eta_1/3q$. 
Comparing the two expressions gives
\begin{equation}
\eta_1={q\eta_c^2\over 6\sqrt 2}\,.\label{eta_1}
\end{equation}
Equations~(\ref{eta_1}) and~(\ref{eta}) provide the leading order form of 
the front bifurcation diagram
\[ c^2={6\sqrt 2\over q\eta_c^2}(\eta_c-\eta)\,,\]
which coincides for small $c$ with the earlier result [6,9],
$c^2=4q^2(\eta_c^2-\eta^2)/\eta^2$.

Multiplying~(\ref{comp2}) by $c$ and using the 
expansions~(\ref{vbar}) and~(\ref{eta}) gives the equation of motion for
propagating fronts
\begin{mathletters}
\label{order}
\begin{eqnarray}
\dot v_f&=
&{\sqrt 2\over q\eta_c^2 }(\eta_c-\eta) 
v_f-{3\over 4\eta_c^2 }v_f^3 \,,\label{ordervf}\\
\dot x_f&=
&-{3\over \eta\sqrt 2} v_f \,, \label{orderyf}
\end{eqnarray}\end{mathletters}where the slow time derivative
of $v_f$ is expressed in terms of a fast time 
derivative ($\dot v_f=c^2 {v_f}_T$).

According to Eqs.~(\ref{order}) the dynamics of a propagating front 
involve two degrees of freedom: a translational degree of freedom, $x_f$, 
determining the front position, and an order parameter, $v_f$, 
determining the direction of propagation. The latter has not been
appreciated enough since most works to date~\cite{TyKe:88,Meron:92}
 have focused on conditions
far from the front bifurcation.  In that case the two stationary states, 
$v_f=\pm\left[4\sqrt 2(\eta_c-\eta)/3q\right]^{-1/2}$, representing fronts
propagating in opposite directions, are highly stable.
Close to the front bifurcation, however, the eigenvalue associated with
these states, $\lambda=-2\sqrt 2(\eta_c-\eta)/q$, approaches zero and 
small disturbances can drive the system from 
one state to another, thereby inducing {\it front reversals}.

%
%
\section{Front reversal:\\ oscillations and rebound }
\label{sec:reversal}
Equations~(\ref{order}) describe the motion of a freely propagating 
front in a uniform medium.
In this Section we show how two different 
perturbations  affect front propagation.
The first is the addition of a space dependent advective field 
to the $v$ equation in the original system (\ref{pde1}).
This type of differential advection appears for example in
chemical reactions involving ionic species that are subjected to
electric fields~\cite{SSM:92,TMPG:94}.
The second is the intrinsic perturbation of uniform curvature 
variations on the propagation of two dimensional fronts.
Both perturbations lead to a coupling of the two degrees of
freedom in the order parameter Eqs.~(\ref{order}) and
allow for the nonsteady asymptotic motion of fronts.

\subsection{Space dependent advective field}
To study the effect of an external 
advective field $J$ we add the term $Jv_x$ to
the $v$ equation in~(\ref{pde1}),
\begin{eqnarray}
\label{pde2}
u_t&=&\epsilon^{-1}(u-u^3-v)+\delta^{-1} u_{xx}\,, \\
 v_t&=&u-a_1v -a_0 + Jv_x+v_{xx}\,. \nonumber
\end{eqnarray}
The small parameter $a_0$ is also introduced to break the
parity symmetry of~(\ref{pde1}).  For simplicity we
consider a linear spatial profile, $J=-\alpha x$, $0<\alpha\ll
1$. Proceeding as before, the inner region analysis remains
unchanged and culminates in~(\ref{orderyf}). The outer analysis
leads to the additional terms, $-\alpha(r+x_f)v_r-a_0$, on the
right hand side of both partial differential equations in~(\ref{free}). 
Assuming $\alpha=\alpha_0 c^3$ and $a_0=a_{00} c^3$,
where $\alpha_0$ and $a_{00}$ are of order unity, Eqs.~(\ref{rho1}) 
and~(\ref{rho2}) remain unchanged, but~(\ref{rho3})
acquires on the right hand side two additional terms:
$\alpha_0(r+x_f)v_r^{(0)} + a_{00}$. As a result the order parameter 
equations~(\ref{order}) are changed to
\begin{mathletters}
\label{order1}
\begin{eqnarray}
\dot v_f&= & {\sqrt 2\over q\eta_c^2 }(\eta_c-\eta)
v_f-{3\over 4\eta_c^2 }v_f^3 +{2\over 3q}\alpha x_f -{4\over
3}a_0 \,, \label{ordervf1}\\ 
\dot x_f&= &-{3\over \eta\sqrt 2} v_f\,. \label{orderyf1}
\end{eqnarray}\end{mathletters}

Notice that the introduction of a {\em space dependent} advective
field couples the two degrees of freedom, $v_f$ and $x_f$. This
coupling affects the front behavior in two significant ways: for
$\eta>\eta_c$ (and $a_0\ne 0$) it stabilizes a propagating front
at a fixed position, $x_f=2qa_0/\alpha$, and for
$\eta<\eta_c$ it induces oscillations between the
counterpropagating fronts. The frequency of oscillations close to
the Hopf bifurcation at $\eta=\eta_c$ is
\begin{equation}
\omega={2\over\sqrt 3}q\sqrt\alpha\,.
\label{freq}
\end{equation}

\begin{figure}
\epsfxsize=3.5 truein \centerline{\epsffile{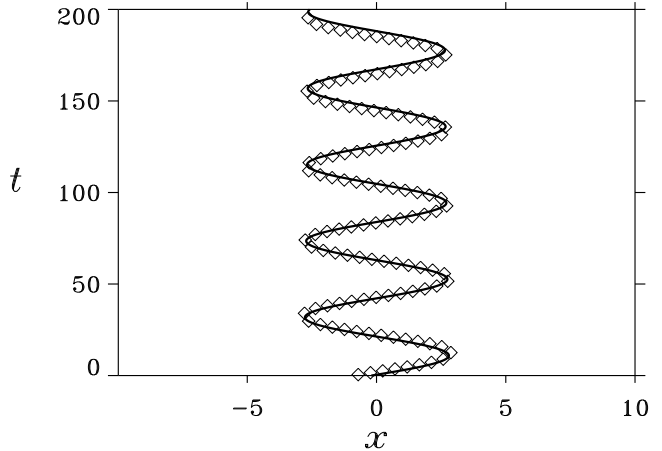}}
\vspace{0.2in}
\caption{
Front position, $x_f$, vs time for an oscillating front.  The
solid line represents the solution to the order parameter equations 
(\protect\ref{order1}) and the diamonds
are from the numerical solution of the original partial differential equations
(\protect\ref{pde1}). 
Parameters: $a_1=3.0,~\epsilon=0.01,~\delta=2.77,~a_0=0$,~$\alpha=0.005$.}
\label{fig:advect}
\end{figure}

\begin{figure}
\epsfxsize=3.5 truein \centerline{\epsffile{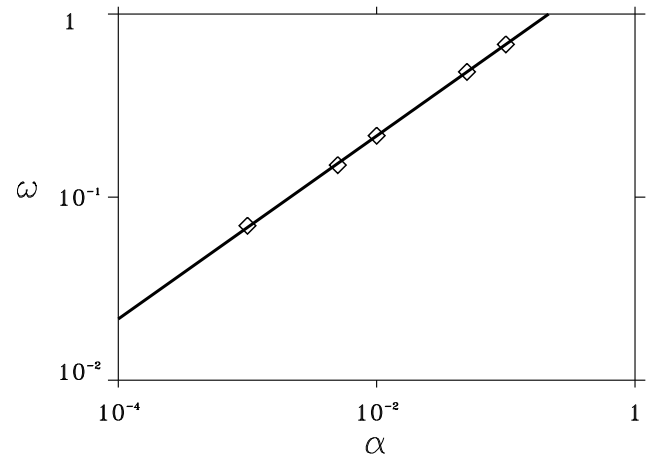}}
\vspace{0.2in}
\caption{
A log-log plot of the oscillation frequency, $\omega$, 
vs the external field
gradient, $\alpha$. The solid line is the relation of
Eq.~(\protect\ref{freq}) and
the diamonds represent numerical solutions of Eqs.~(\protect\ref{pde1}). 
Parameters: $a_1=3.0,~\epsilon=0.01,~\delta=2.77,~a_0=0$.}
\label{fig:omega}
\end{figure}

We tested the validity of Eqs.~(\ref{order1}) by numerically
integrating the original system~(\ref{pde1}) and comparing
the oscillating front solutions with those of~(\ref{order1}). 
The agreement as Fig.~\ref{fig:advect} shows is very
good. In Fig.~\ref{fig:omega} we plotted the frequency of front
oscillations vs the field gradient according to~(\ref{freq}) and
as obtained from~(\ref{pde1}). Again, the agreement is excellent,
and remains good even for $c$ of order unity.
Note that in the inner analysis we neglected contributions of
${\cal O}(\epsilon^2)$ to $v$ 
at the front position, while in the outer analysis
we kept terms to ${\cal O}(c^3)$. A quantitative comparison as described above
therefore requires that $c$ is much larger than $\epsilon^{2/3}$.

\subsection{Uniform curvature}
In two space dimensions the reaction-diffusion system~(\ref{pde1}) becomes
\begin{eqnarray}
\label{pde2d}
u_t&=&\epsilon^{-1}(u-u^3-v)+\delta^{-1}\nabla^2 u\,,\\
v_t&=&u-a_1v-a_0+\nabla^2 v\,, \nonumber
\end{eqnarray}
where the small parameter $a_0$ has been
added again to break the parity symmetry of~(\ref{pde1}).  In
addition to planar front solutions there are now new types of
solutions including fronts with uniform curvature (circular
fronts or spots).  These spots may be stationary or, for
parameters near a NIB bifurcation, may collapse,
expand indefinitely, or oscillate periodically in time.

To derive equations for the motion of circular fronts, the first
step is to transform into polar coordinates,
$r=\rho-\rho_f(t)$, that move with the front.  In this frame
and assuming the radius of curvature $\rho_f$ is much larger
than the front width, Eqs.~(\ref{pde2d}) are
\begin{eqnarray}
u_t-(\dot\rho_f+\delta^{-1}\kappa) u_r&=&
\epsilon^{-1}(u-u^3-v)+\delta^{-1}u_{rr}\,, \nonumber \\
v_t-(\dot\rho_f+\kappa) v_r&=&u-a_1v-a_0+ v_{rr}\,,\nonumber
\end{eqnarray}
where $\kappa=\rho_f^{-1}$ is the front curvature.
As before we assume 
$\epsilon\ll 1$ and $\delta\propto\epsilon^{-1}$ 
and use singular perturbation theory. 

Analysis of the inner, or front, region yields a relation
analogous to~(\ref{orderyf})
\begin{equation}
\dot\rho_f+\delta^{-1}\kappa=-\frac{3}{\eta\sqrt{2}}v_f\,.
\label{rhodot}
\end{equation}
In the outer region, instead of~(\ref{free}) we must solve
\[
v_t+q^2v-v_{rr}=\pm 1-{3\over \eta\sqrt 2}v(0,t)v_r + P\,, 
\]
where $P=(1-\delta^{-1})v_r/\rho_f -a_0$. Assuming that $P$ is a small
perturbation (of order $\vert c\vert ^3$) and proceeding as in 
Section~\ref{solfree} we obtain the order parameter equation
\begin{equation}
\dot v_f={\sqrt 2\over q\eta_c^2 }(\eta_c-\eta) v_f-{3\over
4\eta_c^2 }v_f^3-\frac{2}{3q}\frac{(1-\delta^{-1})}
{\rho_f}- \frac{4}{3}a_0\,.
\label{eq:vf}
\end{equation}

Writing Eqs.~(\ref{rhodot}) and~(\ref{eq:vf}) terms of
the curvature $\kappa=\rho_f^{-1}$ gives the equations
\begin{mathletters}
\label{eq:vf-k}
\begin{eqnarray}
\dot v_f&=&{\sqrt 2\over q\eta_c^2 }(\eta_c-\eta) v_f-{3\over
4\eta_c^2 }v_f^3-\frac{2}{3q}(1-\delta^{-1})\kappa-
\frac{4}{3}a_0\,, \label{order2} \\
\dot\kappa&=&\frac{3}{\eta\sqrt
2}v_f\kappa^2+\delta^{-1}\kappa^3\,,
\label{kappadot}
\end{eqnarray}
\end{mathletters}that describe the dynamics of large circular spots.  
The introduction of curvature {\em couples} the two
equations.  Equations~(\ref{order}) for planar fronts are
decoupled and only describe the relaxation to steadily
propagating fronts.  The equations for circular fronts
additionally allow front reversals and
nonsteady asymptotic motion like oscillations.

\begin{figure}
\epsfxsize=3.5 truein \centerline{\epsffile{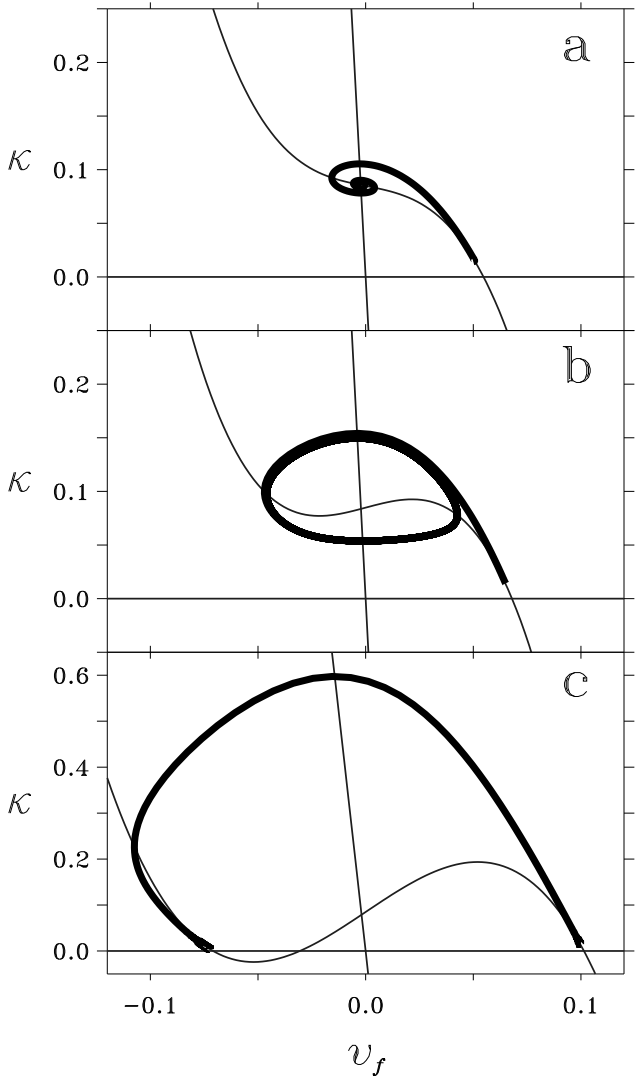}}
\vspace{0.6in}
\caption{Three types of solutions to the order parameter
equations (\protect\ref{eq:vf-k}) starting with initial
conditions representing a large shrinking spot.
The thin lines are the isoclines and the thick lines are
numerically computed trajectories.
(a) Convergence to a stationary spot ($\epsilon=.0063$).
(b) An oscillating spot ($\epsilon=.006$).
(c) Spot rebound and expansion of the spot to infinite size ($\epsilon=.0052$).
Parameters:  $a_1=4.0$, $a_0=-0.01$, and $\delta=2.0$.}
\label{fig:odes}
\end{figure}

Consider first the fixed point solutions obtained by the
intersections of the linear nullclines $\kappa=0$ and
$\kappa=-(3\delta/\eta\sqrt 2)v_f$ of~(\ref{kappadot})
with the cubic nullcline of~(\ref{order2}).  The solutions
corresponding to $\kappa=0$ describe planar fronts propagating
at constant velocities.  Solutions with positive and negative
$v_f$ values pertain to down states invading up states and up
states invading down states, respectively. The number of
$\kappa=0$ solutions varies with $\eta$.  Below the front
bifurcation, [$\eta>\eta_c(a_0)$], there is a single
intersection point representing an Ising front as shown by the
thin lines in
Figs.~\ref{fig:odes}a and~\ref{fig:odes}b. Beyond the front
bifurcation, [$\eta<\eta_c(a_0)$], two more intersection
points appear corresponding a stable and unstable pair of
planar front solutions (Fig.~\ref{fig:odes}c).  The fixed
point solutions for $\kappa\neq 0$
represent a circular fronts.  For $a_0<0$ they describe spots
of up an state domain and for $a_0>0$ spots of a down state domain.
For $\delta>1$, depending on the choice of $\epsilon$, these
fixed points may or may not be stable.  For $\delta<1$, all
the $\kappa \neq 0$ fixed points are {\it unstable}.

\begin{figure}
\epsfxsize=3.5 truein \centerline{\epsffile{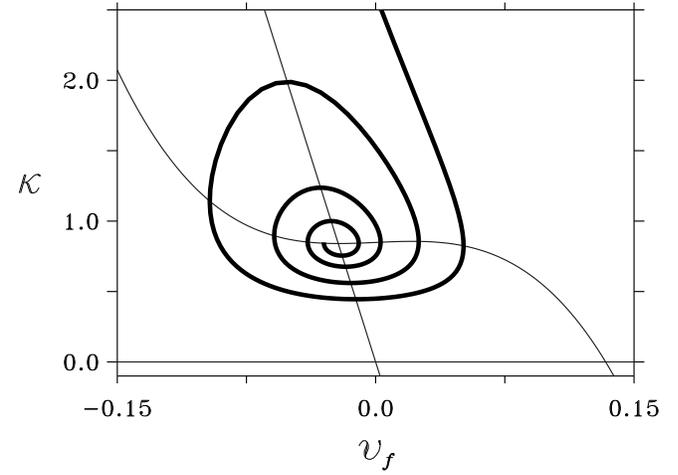}}
\vspace{0.4in}
\caption{A trajectory of the order parameter equations (\protect\ref{eq:vf-k})
for a spot that oscillates with growing
amplitude until collapse (the curvature $\kappa$ diverges to infinity). 
Parameters:  $a_1=4.0$, $a_0=-0.1$, $\epsilon=0.006$, and $\delta=2.0$.}
\label{fig:collapse}
\end{figure}

Figure~\ref{fig:odes} shows three different possibilities for
the dynamics of circular fronts.  The thick trajectories
represent dynamics computed by numerical solution of the
coupled equations~(\ref{eq:vf-k}).  The initial conditions
correspond to a large shrinking up state spot. Far into the
Ising regime (Fig.~\ref{fig:odes}a) the initial spot converges
to a stationary spot.  Moving closer to the front bifurcation
and past a critical $\eta$ value, $\eta_H>\eta_c(a_0)$, a Hopf
bifurcation to a breathing spot occurs (Fig.~\ref{fig:odes}b).
Crossing the front bifurcation, $\eta<\eta_c(a_0)$, the spot
rebounds, i.e. the shrinking spot reaches a minimal size and
expands again indefinitely (Fig.~\ref{fig:odes}c).  For larger
$\vert a_0\vert$, there is another possibility for the dynamics
of spots.  In this case, shown in Fig.~\ref{fig:collapse}, the
amplitude of oscillations grows in time until the spot
eventually collapses as the curvature diverges to infinity.

\begin{figure}
\epsfxsize=3.5 truein \centerline{\epsffile{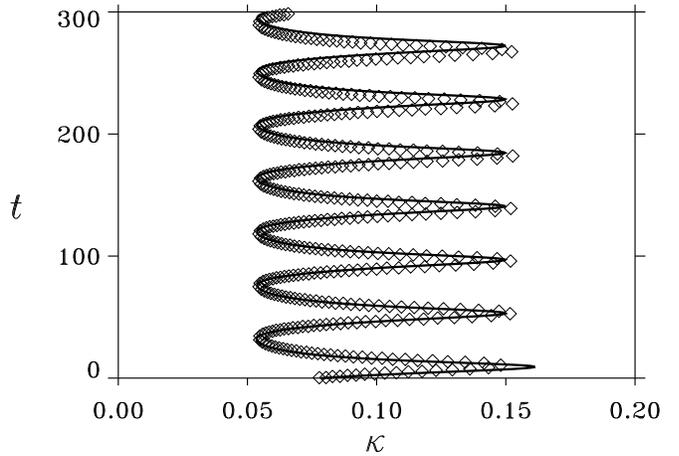}}
\vspace{0.2in}
\caption{An oscillating circular spot solution.  The solid line is the 
solution 
of the order parameter equations (\protect\ref{eq:vf-k}),  
and the diamonds
represent the spot curvature vs time from the 
numerical solution to the circularly symmetric equations 
(\protect\ref{eq:circular}). The
equation parameters are $\epsilon=0.006, \delta=2.0, a_1=4.0,a_0=-0.01.$ }
\label{fig:compare}
\end{figure}

All the three behaviors discussed above have been observed
in direct numerical solutions of~(\ref{pde2d}).
The quantitative accuracy of the order parameter equations
was tested by computing numerical solutions to 
the circularly symmetric version of Eqs.~(\ref{pde2d})
\begin{eqnarray}
\label{eq:circular}
u_t&=& \epsilon^{-1}(u-u^3-v)+{\delta^{-1}\over r}u_r+\delta^{-1}u_{rr}\,,\\
v_t&=&u-a_1v-a_0+ \frac{1}{r}v_r+v_{rr}\,. \nonumber
\end{eqnarray}
and comparing them to solutions to Eqs.~(\ref{eq:vf-k}) for
spot dynamics.  Spot solutions of~(\ref{eq:circular}) produce the
same qualitative behavior as the pair of coupled equations for
the spot dynamics.  When the parameters are chosen to satisfy the
assumptions made in the derivation of~(\ref{eq:vf-k}), there is
also quantitative agreement between the two solutions.
Figure~\ref{fig:compare} shows the curvature of an oscillating
spot as a function of time computed using both Eqs.~(\ref{eq:vf-k})
and~(\ref{eq:circular}).  
The two solutions agree within an accuracy of approximately 1\% for
the amplitude and 2\% for the phase.

In addition to the oscillatory instability spot solutions may also
be unstable to transverse perturbations~\cite{OMK:89,PeGo:94,GMP:96}.  
Numerical solutions of the fully
two-dimensional model~(\ref{pde2d}) show that for the parameters of
Fig.~\ref{fig:compare} spots are unstable and form 
nonuniformly curved fronts leading to a labyrinthine pattern.
Since the
order parameter equations derived here apply only for the case when
the spots do not break perfect circular symmetry, for this
choice of parameters they only capture
the dynamics of the circular spot during the initial evolution.
Order parameter equations for the dynamics of nonuniformly curved
fronts are presented in Ref.~\cite{HaMe:97}.

%
%
\section{Conclusion}

We derived the equations that govern the dynamics of planar
fronts in bistable systems near a parity breaking front bifurcation
(the NIB bifurcation).  In this case the context is an
activator-inhibitor model, but the normal form
\begin{eqnarray}
\label{normal}
\dot X &=& C\,,  \\ 
\dot C &=& (\alpha_c-\alpha)C-\beta C^3\,, \nonumber
\end{eqnarray}

is general.  Here, $X$ is the front position, $C$ is the front
velocity, and $\alpha_c$ is a critical parameter value for
which a NIB bifurcation occurs.  Similar
equations should apply, for example, to liquid crystals
subjected to rotating magnetic fields~\cite{NYK:95,FRCG:94,MiMe:94}
and have also been proposed in the context
of parity breaking traveling-wave bifurcations~\cite{KHD:95}.

Uniform front curvature, or space dependent external fields, couple the
two degrees of freedom, $X$ and $C$, and allow nonsteady asymptotic
behavior.  The coupled sets of equations~(\ref{order1})
and~(\ref{eq:vf-k}) exhibit Hopf bifurcations from stationary to
oscillating fronts (breathing spots).  Equations~(\ref{eq:vf-k}) exhibit
two additional behaviors pertaining to rebounding and collapsing spots in
the full equations. 

Curvature effects on front dynamics near a NIB bifurcation were also
studied in Refs.~\cite{EHM:95,HaMe:94c} using a ``quasistatic''
approximation.  This approximation, where the front velocity is assumed to
adiabatically follow slow curvature variations~\cite{TyKe:88,Meron:92},
yields an {\it algebraic} relation between the front velocity $C$ and its
curvature $\kappa$.  As the bifurcation is approached the $C-\kappa$
relation becomes multivalued, or hysteretic.  The multivalued relations
correctly predict spontaneous front transitions induced by
curvature~\cite{EHM:95,HaMe:94c}, but cannot describe dynamics during
front transitions.  Differential order parameter equations,
like~(\ref{eq:vf-k}), give a more accurate characterization of the
dynamics.  These differential equations reduce to an algebraic $C-\kappa$
relation when the time scale of front transitions becomes much shorter
than the time scale of curvature changes.  Such a condition is realized,
for example, with very large spots away from the front bifurcation.  Then
the right hand side of~(\ref{order2}) can be set to zero, an expression
which together with~(\ref{rhodot}) gives an
algebraic $C-\kappa$ relation where $C=\dot\rho_f$.

The phenomena of breathing, oscillating, and collapsing spots appear to be
quite general and can be induced by other perturbations that couple $X$
and $C$.  Ref.~\cite{Haim:96}, which studies the effect of boundaries on
spot dynamics, reports on the observation of stationary, breathing, and
rebounding spots.  Interaction between fronts may similarly lead to
stationary, oscillating, and collapsing
domains~\cite{KoKu:80,KeOS:82,NiMi:89,OITe:90,RRHM:93,HaMe:94a,SOMS:95,MuOs:96,WSBOP:96}. Recent experiments on the Ferrocyanide-Iodate-Sulfite
reaction show small oscillating chemical spots away from the reactor
boundary that are most likely due to front interactions and/or
curvature~\cite{LOS:96}.

\acknowledgments
We thank the anonymous referee for suggesting a shorter method
for the solution of Eqs.~(\ref{prob}).
This study was supported in part by the Israel Ministry of Science and the 
Arts.

\bibliography{reaction}

\begin{thebibliography}{10}

\bibitem{TyKe:88}
J.~J. Tyson and J.~P. Keener, Physica D {\bf 32},  327  (1988).

\bibitem{Meron:92}
E. Meron, Physics Reports {\bf 218},  1  (1992).

\bibitem{EMS:88}
C. Elphick, E. Meron, and E.~A. Spiegel, Phys. Rev. Lett. {\bf 61},  496
  (1988).

\bibitem{EMS:90}
C. Elphick, E. Meron, and E.~A. Spiegel, SIAM J. Appl. Math {\bf 50},  490
  (1990).

\bibitem{AKP:84}
K.~I. Agladze, V.~I. Krinsky, and A.~M. Pertsov, Nature {\bf 308},  834
  (1984).

\bibitem{MPH:88}
S.~C. M\"uller, T. Plesser, and B. Hess,  in {\em Physiochemical Hydrodynamics:
  Interfacial Phenomena}, edited by M.~G. Velarde (Plenum Press, New York,
  1988).

\bibitem{Ort:87}
P. Ortoleva, Physica D {\bf 26},  67  (1987).

\bibitem{SSM:92}
O. Steinbock, J. Sch\"utze, and S.~C. M\"uller, Phys. Rev. Lett. {\bf 68},  248
   (1992).

\bibitem{MHM:94}
A.~F. Munster, P. Hasal, and M. Marek, Phys. Rev. E {\bf 50},  546  (1994).

\bibitem{KoKu:80}
S. Koga and Y. Kuramoto, Prog. Theor. Phys. {\bf 63},  106  (1980).

\bibitem{KeOS:82}
B.~S. Kerner and V.~V. Osipov, Sov. Phys. JETP {\bf 56},  1275  (1982).

\bibitem{NiMi:89}
Y. Nishiura and M. Mimura, SIAM J. Appl. Math. {\bf 49},  481  (1989).

\bibitem{OITe:90}
T. Ohta, A. Ito, and A. Tetsuka, Phys. Rev. A {\bf 42},  3225  (1990).

\bibitem{RRHM:93}
Y.~A. Rzanov, H. Richardson, A.~A. Hagberg, and J.~V. Moloney, Phys. Rev. A
  {\bf 47},  1480  (1993).

\bibitem{HaMe:94a}
A. Hagberg and E. Meron, Nonlinearity {\bf 7},  805  (1994).

\bibitem{SOMS:95}
M. Suzuki, T. Ohta, M. Mimura, and H. Sakaguchi, Phys. Rev. E {\bf 52},  3645
  (1995).

\bibitem{MuOs:96}
C.~B. Muratov and V.~V. Osipov, Phys. Rev. E {\bf 53},  3101  (1996).

\bibitem{WSBOP:96}
R. Woesler, P. Sch\"utz, M. Bode, M. Or-Guil, and H.-G. Purwins, Physica D {\bf
  91},  376  (1996).

\bibitem{HaMe:94b}
A. Hagberg and E. Meron, Phys. Rev. Lett. {\bf 72},  2494  (1994).

\bibitem{EHM:95}
C. Elphick, A. Hagberg, and E. Meron, Phys. Rev. E {\bf 51},  3052  (1995).

\bibitem{LMOS:93}
K.~J. Lee, W.~D. McCormick, Q. Ouyang, and H.~L. Swinney, Science {\bf 261},
  192  (1993).

\bibitem{LMSP:94}
K.~J. Lee, W.~D. McCormick, J.~E. Pearson, and H.~L. Swinney, Nature {\bf 369},
   215  (1994).

\bibitem{LeSw:95}
K.~J. Lee and H.~L. Swinney, Phys. Rev. E {\bf 51},  1899  (1995).

\bibitem{CLHL:90}
P. Coullet, J. Lega, B. Houchmanzadeh, and J. Lajzerowicz, Phys. Rev. Lett.
  {\bf 65},  1352  (1990).

\bibitem{HaMe:94c}
A. Hagberg and E. Meron, Chaos {\bf 4},  477  (1994).

\bibitem{Bode:96}
M. Bode, ``{F}ront bifurcations in reaction-diffusion systems with
  inhomogeneous parameters'' Submitted to Physica D (1996) (unpublished).

\bibitem{HaMe:97}
A. Hagberg and E. Meron, ``{T}he dynamics of curved fronts: Beyond geometry'',
  to appear in {Physical Review Letters} (unpublished).

\bibitem{HMRZ:96}
A. Hagberg, E. Meron, I. Rubinstein, and B. Zaltzman, Phys. Rev. Lett. {\bf
  76},  427  (1996).

\bibitem{BNREG:95}
M. B\"ar, S. Nettesheim, H.~H. Rotermund, M. Eiswirth, and G.Ertl, Phys Rev.
  Lett. {\bf 74},  1246  (1995).

\bibitem{HBKR:95}
G. Haas, M. B\"ar, I.~G. Kevrekidis, P.~B. Rasmussen, H.-H. Rotermund, and
  G.Ertl, Phys Rev. Lett. {\bf 75},  3560  (1995).

\bibitem{BRSP:94}
M. Bode, A. Reuter, R. Schmeling, and H.-G. Purwins, Phys Lett. A {\bf 185},
  70  (1994).

\bibitem{SBP:95}
P. Sch\"utz, M. Bode, and H.-G. Purwins, Physica {\bf 82D},  382  (1995).

\bibitem{IMN:89}
H. Ikeda, M. Mimura, and Y. Nishiura, Nonl. Anal. TMA {\bf 13},  507  (1989).

\bibitem{TMPG:94}
J.~J. Taboada, A.~P. {Mu\~{n}uzuri}, V. {P\'{e}rez-Mu\~{n}uzuri}, M.
  G\'{o}mez-Gesteira, and V. P\'{e}rez-Villar, Chaos {\bf 4},  519  (1994).

\bibitem{OMK:89}
T. Ohta, M. Mimura, and R. Kobayashi, Physica D {\bf 34},  115  (1989).

\bibitem{PeGo:94}
D.~M. Petrich and R.~E. Goldstein, Phys. Rev. Lett. {\bf 72},  1120  (1994).

\bibitem{GMP:96}
R.~E. Goldstein, D.~J. Muraki, and D.~M. Petrich, Phys. Rev. E {\bf 53},  3933
  (1996).

\bibitem{NYK:95}
S. Nasuno, N. Yoshimo, and S. Kai, Phys. Rev. E {\bf 51},  1598  (1995).

\bibitem{FRCG:94}
T. Frisch, S. Rica, P. Coullet, and J.~M. Gilli, Phys. Rev. Lett. {\bf 72},
  1471  (1994).

\bibitem{MiMe:94}
K.~B. Migler and R.~B. Meyer, Physica D {\bf 71},  412  (1994).

\bibitem{KHD:95}
E. Knobloch, J. Hettel, and G. Danglemayr, Phys. Rev. Lett. {\bf 74},  4839
  (1995).

\bibitem{Haim:96}
D. Haim, G. Li, Q. Ouyang, W.~D. McCormick, H.~L. Swinney, A. Hagberg, and E.
  Meron, Phys. Rev. Lett. {\bf 77},  190  (1996).

\bibitem{LOS:96}
G. Li, Q. Ouyang, and H.~L. Swinney, J. Chem. Phys {\bf 105},  10830  (1996).

\end{thebibliography}

\end{document}